\documentclass[aps,prper,showpacs,twocolumn,floats,superscriptaddress,10pt]{revtex4-1}
\usepackage{graphicx}
\usepackage{dcolumn}
\usepackage{bm}
\usepackage{amssymb}
\usepackage{nicefrac}
\usepackage{appendix}
\usepackage{amsmath}

\usepackage[latin3]{inputenc}
\usepackage[makeroom]{cancel}
\usepackage{amsmath}
\usepackage{amsfonts}
\usepackage{amssymb}
\usepackage{color}
\usepackage[left=2cm,right=2cm,top=2cm,bottom=2cm]{geometry}

\usepackage{graphicx} 
\usepackage{dcolumn}
\usepackage{bm}
\usepackage{simplewick}
\usepackage{array}
\usepackage{appendix}

\RequirePackage[
   hyperindex,colorlinks,bookmarksnumbered,
   plainpages=true,pdfstartview=FitH]{hyperref}
\hypersetup{linkcolor=blue,urlcolor=blue,citecolor=blue}
\usepackage{hyperref}

\definecolor{purple}{rgb}{0.5,0,0.6}

\usepackage{ulem}

\begin{document}

\title{Generalized Cutler-Mott relation in a two-site charge Kondo simulator}

\date{\today}

\author{T. K. T. Nguyen}
\email{nkthanh@iop.vast.vn}
\affiliation{Institute of Physics, Vietnam Academy of Science and Technology, 10 Dao Tan, 118000 Hanoi, Vietnam}
\author{M. N. Kiselev}
\affiliation{The Abdus Salam International Centre for Theoretical Physics, Strada
Costiera 11, I-34151, Trieste, Italy}

\begin{abstract}
We analyze the validity of the Cutler-Mott relations outside the Landau Fermi-liquid concept. We consider a 
two-site charge Kondo circuit as a paradigmatic example of a system possessing both Fermi- and  non-Fermi liquid properties. It is shown that the
generalized Cutler-Mott-like relations derived in the paper hold in both operating regimes of the charge Kondo quantum circuit describing a smooth crossover between low- and high- temperature regimes. We discuss applicability of the generalized Cutler-Mott relations for computing a figure of merit of the non-Fermi liquid quantum simulators.
\end{abstract}

\maketitle

\section{Introduction}
Thermoelectricity, which describes the direct conversion between heat and electrical energy, has attracted considerable attention in recent years from both physicists and engineers. The study of thermoelectric effects provides valuable insight into the electronic structure and fundamental scattering processes within a system. Furthermore, thermoelectric materials can be utilized to generate electricity, measure temperature, or alter the temperature of objects due to the reversible effects of Seebeck, Peltier, and Thomson \cite{Wood}.

The Seebeck effect \cite{Seebeck} happens when a temperature gradient is applied across two dissimilar electrical conductors or semiconductors, resulting in a voltage difference between them \cite{TEbook1,TEbook2}. The efficiency of this conversion of thermoelectric energy is characterized by the Seebeck coefficient, or thermopower (TP), which is defined as the ratio of the generated electric voltage $V_{th}$ to the temperature difference $\Delta T$: $S=-eV_{th}/\Delta T$. The measurement of the thermovoltage, $V_{th}$, offers independent information about the thermoelectric coefficient $G_T$. The temperature difference $\Delta T$, is regulated by using a current heating technique. The differential electric conductance $G$ is measured at a variable external energy source. In the linear regime, the TP is typically determined by the relation $S=G_T/G$. TP is a fundamental property that defines the thermoelectric performance of a material, making its enhancement both crucial and challenging. With advances in nanotechnology, numerous nanostructured devices have demonstrated promising improvements in TP.

In low-dimensional systems, the TP is expected to be greater than in three-dimensional systems for a given carrier concentration because of size-quantization effects. To understand this, we refer to the Cutler-Mott (CM) relation for TP. This relation, introduced by Cutler and Mott in 1969 \cite{CM}, is a fundamental theoretical framework for understanding thermoelectric phenomena in metallic and semiconducting materials. It presents a key connection between the TP and the electrical conductivity $G$,
\begin{equation}
S=\frac{\pi^2}{3}\frac{k_B^2 T}{e}\left[\frac{d\ln[G(E)]}{dE}\right]_{E=E_F},
\label{CM}
\end{equation}
providing deep insight into the electronic structure and scattering mechanisms of a material. It has become a cornerstone in the study of thermoelectric properties, particularly in the context of metallic and metallic-like materials, where it describes the temperature dependence of the TP.

The CM relation is a crucial aspect to examine when investigating thermoelectric materials. As shown in Eq. (\ref{CM}), it exhibits several notable properties \cite{dollfus}. First, the formula includes a term $k_B/e \sim 80\mu V/K$, which can be regarded as the natural unit of TP. Second, the derivative of the logarithm of conductivity in the expression highlights that TP is a highly sensitive probe of the electronic structure of the material. Third, the linear temperature dependence observed suggests that this behavior is characteristic of metallic and metallic-like materials. Finally, the CM formula encodes information about scattering processes within the material.

It is well-established that the TP is a valuable measure of Kondo correlations \cite{Kondo_TP}. Thermodynamically, it is consistent with the entropy associated with the flow of charge in a material. In typical metals and degenerate semiconductors, only a fixed population of electrons, those with energies close to the Fermi energy within a few temperatures, contribute to both charge and thermoelectric transport. As the temperature decreases, the narrowing of the Fermi-Dirac distribution leads to a linear reduction in the Seebeck coefficient. Under the rigid band approximation, the CM formula is commonly used to interpret the TP. This observation has prompted us to investigate the validity of the CM relation in charge Kondo systems.

The charge Kondo effect is a Kondo-like phenomenon that maps the degrees of freedom of the system onto those of the conventional Kondo effect. Specifically, the charge Kondo effect involves the degeneracy of charge states (for instance, $N$ and $N+1$ electron states) at the Coulomb peaks in a quantum dot (QD). These two charge states can be mapped to the two spin projections of a spin-$1/2$ magnetic impurity in the conventional Kondo effect, allowing the QD to be treated as an iso-spin-$1/2$ magnetic impurity. The charge Kondo model was first proposed by Flensberg, Matveev, and Furusaki (FMF) \cite{flensberg,matveev,furusakimatveev}, in which a large metallic QD is strongly coupled to one or more leads through one or several almost transparent quantum point contacts (QPCs).

An alternative interpretation of the charge Kondo phenomenon in the FMF model is that electrons inside the QD are assigned to the ``iso-spin up" state, while those outside the QD are assigned to the ``iso-spin down" state. The iso-spin flips when electrons move in and out of the QD. Backscattering at the QPC transfers electrons from the "moving in" state to the ``moving out" state and vice versa. The additional internal degrees of freedom, such as the spin projection quantum number of electrons or the number of single-mode QPCs, determine the number of distinct channels in the Kondo problem \cite{flensberg,matveev,furusakimatveev,MAprl,MAprb,lehur,thanh2010,thanh2015}. The charge Kondo effect has recently been observed in the integer quantum Hall (IQH) regime through a series of breakthrough experiments \cite{pierre2,pierre3}, which expand the experimental possibilities for accessing multi-channel Kondo (MCK) physics (three channels or more).

It is well established that deviations from the CM law in thermoelectric transport through QD systems in the conventional Kondo regime originate from strong interactions between localized spins and conduction electron spins \cite{exp2005}. Interestingly, in the FMF setup of the two-channel charge Kondo model, the CM relation remains valid when the system resides in the Fermi liquid (FL) regime \cite{MAprl,thanh2010,Karki2020}. Moreover, deviations from the CM law under finite in-plane magnetic fields within the FL regime can serve as a precursor to non-Fermi liquid (NFL) behavior \cite{thanh2010,thanh2015}.

This raises a key question: Can the CM relation be generalized to systems in the NFL state? If so, such a generalization could provide a powerful tool for probing the TP of charge Kondo systems.

In this work, we propose a generalized Cutler-Mott (GCM) relation for a two-site charge Kondo circuit (2SCKC), and demonstrate its applicability to both FL and NFL regimes. Furthermore, we show that the GCM relation can be extended to other charge Kondo setups and used to analyze the thermoelectric figure of merit.

The paper is structured as follows. In Sec. II, we introduce the theoretical model and review the standard CM relation in the context of the 2SCKC. Sec. III presents our proposed GCM relation and discusses its physical implications. In Sec. IV, we explore the extension of GCM to other charge Kondo systems. Sec. V focuses on the application of GCM to the figure of merit. Finally, we summarize our findings and conclusions in Sec. VI.
\begin{figure}[t]
\begin{center}
\includegraphics[angle=-90,scale=0.3]{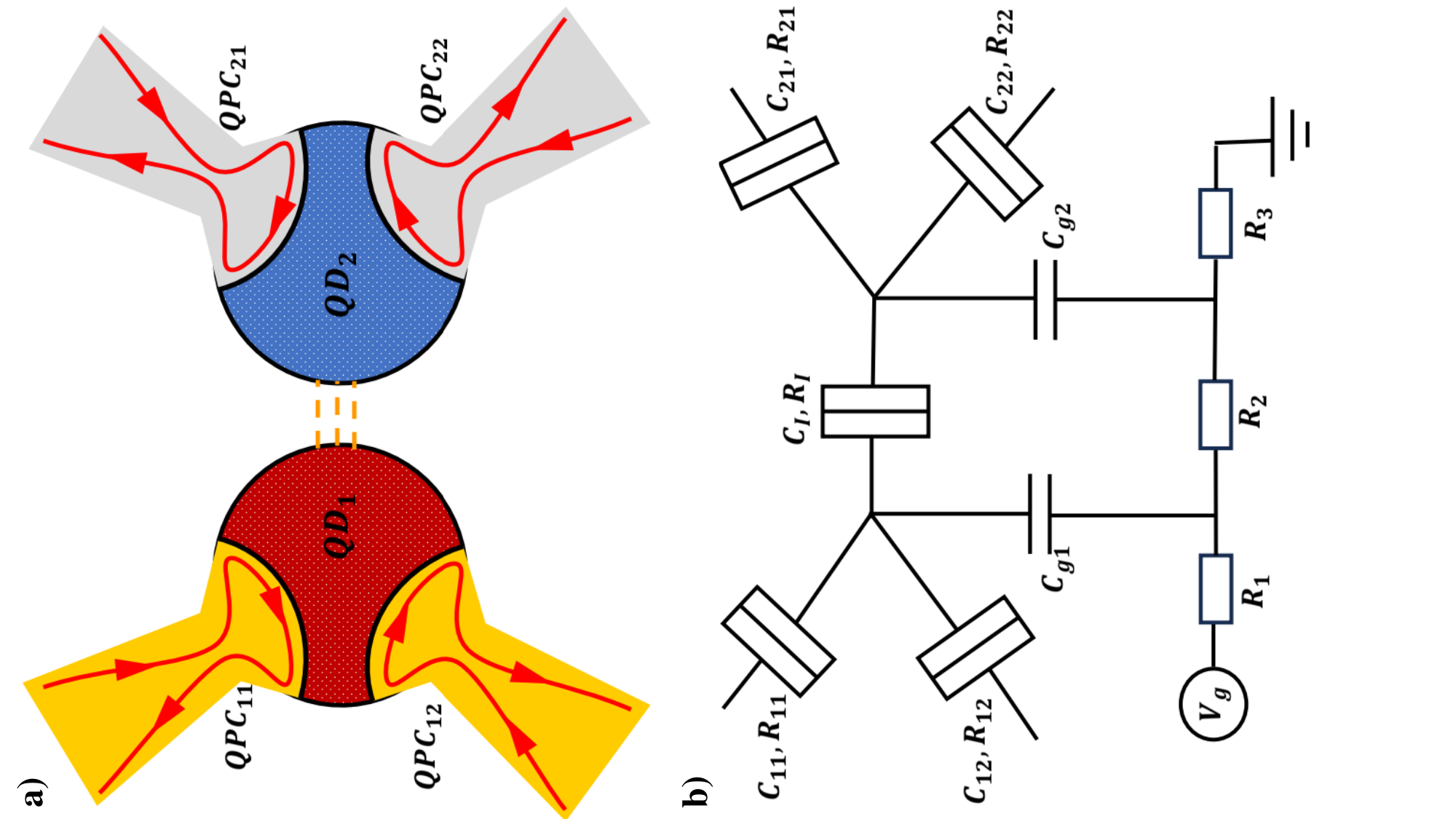}
\end{center}
\vspace{-1.2cm}
\caption{a) Schematic of a weak link between two charge Kondo circuits (CKCs). Each circuit consists of a large metallic island (QD), which is embedded into two-dimensional electron gas (2DEG) and connects to two large electrodes through the single-mode quantum point contacts (QPCs). The 2DEG (plain area) is in the integer quantum Hall regime $\nu=1$. The red line with arrows denotes the chiral edge mode which backscatters at the center of the narrow constriction. The QPCs are fine tuned to different regimes. The right CKC (grey color) is at the reference temperature $T$ while the left circuit (orange color) is at higher temperature $T+\Delta T$. b) The equivalent circuit where the capacitors $C_{g1}$, $C_{g2}$ and the part to support the gate voltages for the QDs are not shown in a). The resistors $R_1$, $R_2$, $R_3$ are used to divide the input gate voltage $V_g$.}
\label{Fig1}
\end{figure}

\section{Model and Cutler-Mott relation of a two-site charge Kondo circuit}

In this paper, we examine the validity of the Cutler-Mott (CM) formula in a two-site charge Kondo (2SCK) simulator. The model, illustrated in Fig. \ref{Fig1}a), was theoretically proposed in Ref. \cite{thanh2018} and has recently been experimentally implemented \cite{Gordon2023}. Each charge Kondo circuit (CKC) consists of a quantum dot (QD)-quantum point contact (QPC) structure. The QD is a large metallic island (depicted in dark red and blue cross-hatched areas, surrounded by black lines) that is electronically connected to a two-dimensional electron gas (2DEG, represented by the orange and grey continuous areas). The 2DEG is coupled to two large electrodes via two QPCs. By applying a strong magnetic field perpendicular to the 2DEG plane, the 2DEG can be controlled in the integer quantum Hall (IQH) regime at the filling factor $\nu=1$, a setup that has been explored experimentally \cite{pierre2,pierre3}. The QPCs are finely tuned to be nearly transparent, meaning that the chiral edge currents (depicted as red solid lines with arrows) undergo weak backscattering at the QPCs. In this work, we investigate a configuration in which the reflection amplitudes are identical at both QPCs in each CKC: $|r_{11}|=|r_{12}|=|r_{1}|$ and $|r_{21}|=|r_{22}|=|r_{2}|$, representing the two-channel Kondo (2CK) setup of each CKC. The left CKC is set to a higher temperature, $T+\Delta T$, compared to the right circuit, which is maintained at a temperature $T$. The temperature difference creates a drop across the central weak link.

The equivalent circuit of the device is shown in Fig. \ref{Fig1} b). The $C_{\alpha,j},R_{\alpha,j}$ (${\alpha,j}=11,\, 12,\, 21,\, 22$) and $C_I,R_I$ characterize the QPCs and the weak central coupling, respectively. Additionally, we include the capacitors $C_{g1}$, $C_{g2}$ and the gate voltages for the quantum dots (QDs), which are not shown in Fig. \ref{Fig1}a. The resistors $R_1$, $R_2$, $R_3$ are used to divide the input gate voltage $V_g$. Consequently, the dimensionless gate voltages of the QDs are estimated as follows:
\begin{equation}
N_1=\frac{C_{g1}}{e}\frac{R_2+R_3}{R_t}V_g,\;  N_2=\frac{C_{g2}}{e}\frac{R_3}{R_t}V_g,
\label{gate}
\end{equation} 
with $R_t=R_1+R_2+R_3$. In fact, the resistors are variable. We find that if $C_{g1}=C_{g2}$ we can have $N_1=N_2$ if $R_2=0$ or $N_2=0$ if $R_3=0$. 

CM relation allows us to compute the TP in the 2SCK circuit as
\begin{equation}
S_{CM}\!=\!\frac{\pi^2}{6e}\!\!\sum_{j=1,2}\!\!\!\mathcal{R}_j\frac{T}{E_{C,j}}\frac{\partial\ln G}{\partial N_j}\!=\!\frac{\pi^2}{6e}\!\!\sum_{j=1,2}\!\!\frac{T}{E_{C,j}}\frac{\partial\ln G}{\partial N_j},
\label{SCM0}
\end{equation}
where $\mathcal{R}_1=R_3/R_t$, $\mathcal{R}_2=( R_2+R_3)/R_t$ are the ratios of the resistances in the voltage divider. These ratios are independent of the 2SCK device. Without loss of generality, we choose $R_1=R_2=0$ so that $\mathcal{R}_1=\mathcal{R}_2=1$. This leads to the second equality.

The thermoelectric coefficients in the 2SCK model within the linear response regime -- where $[\Delta T, e\Delta V] \ll T$ -- have been computed and discussed in detail in Refs.\cite{thanh2018, thanh2024, thanhcom2022}. The general expressions are presented in Appendix \ref{previous}.

For the single-site charge Kondo circuit, it is mentioned that the conventional CM relation holds in the FL regime, where the temperature is much lower than the Kondo resonance width [$T\ll \Gamma$, where $\Gamma(N)=8\gamma E_C|r|^2\cos^2(\pi N)/\pi^2$, with $E_C$ is the charging energy of the QD, $\gamma=e^{\textbf{C}}\approx1.78$ ($\textbf{C}\approx 0.577$ is Euler's constant)] \cite{MAprl, thanh2010, Karki2020}. However, this relation breaks down in the NFL regime, in which $\Gamma\le T\ll E_C$. To address this, we propose a GCM relation that remains valid in both FL and NFL regimes, covering the entire temperature range $T\ll E_C$.

\section{Generalized Cutle-Mott formula}

As highlighted in Ref.\cite{MAprl}, in the low-temperature regime ($T\ll \Gamma$), the system adheres to the FL paradigm, where physical observables exhibit characteristic behavior -- most notably, the conductance follows a quadratic temperature dependence, $G\propto T^2$. In this regime, TP $S$ can be expressed in terms of the logarithmic derivative of conductance with respect to gate voltage, reflecting the sensitivity of the transport to electrostatic tuning.

This form is reminiscent of the CM formula, originally derived for metals with noninteracting electrons, where TP is directly related to the energy derivative of the conductivity. However, in strongly correlated systems, such as those governed by Coulomb interactions, the prefactor in this relation is significantly modified. In particular, it acquires an additional large factor
$\ln(E_C/\Gamma)$, which encapsulates the influence of interaction-driven energy scales.

In the high-temperature regime ($T\gg \Gamma$), the system crosses over into a NFL state, where the simple logarithmic derivative form of TP breaks down, and no direct analog to the CM formula exists. This discontinuity between the FL and NFL regimes highlights a conceptual and practical limitation in the existing framework for understanding thermoelectric response across temperature regimes.

Therefore, to consistently describe TP in both FL and NFL regimes -- and to account for interaction-driven modifications -- a generalized CM relation is necessary. Such a formulation would extend the applicability of the original CM approach, incorporating both the effects of strong correlations and the crossover between different transport regimes.

We propose a novel expression for thermoelectric response in the 2SCK model, referred to as the GCM. It links the TP $S$ to the energy derivative of the electric conductance $G$, incorporating interaction effects through the Kondo resonance widths $\Gamma_j$. This relation is a extension of the original CM formula for nanostructured systems with strong electronic correlations.
\begin{equation}
S_{GCM}=\frac{\pi^2}{6e}\sum_{j=1,2}\frac{T}{E_{C,j}}\ln\left[\frac{E_{C,j}}{T+\Gamma_j}\right]\frac{\partial\ln G}{\partial N_j}.
\label{SCM}
\end{equation}

As shown in the Appendix  \ref{previous}, the analytical expression for the function $F_C(p_1, p_2)$ in Eq.~(\ref{FC}) has been evaluated in Ref.~\cite{Kiselev2025}. Notably, in the limiting cases where either $p_1 = 0$ or $p_2 = 0$, corresponding to $\Gamma_1 = 0$ or $\Gamma_2 = 0$, exact analytical results can be obtained as
\begin{eqnarray}
&& F_{C}\left(0,p_{2}\right)\!\!=\!\!\pi F_{C,0}(p_2), \;\;\;
F_{C}\left(p_1,0\right)\!=\!\pi F_{C,0}(p_1),\nonumber\\
&& F_{C,0}(x)\!\!=\!\!\pi\!\left(4\pi^2-x^2\right)\!\left[1-\frac{2x}{\pi}\Phi\!\left(-1,1,1+\frac{x}{\pi}\!\right)\!\right]\nonumber\\
&&+\frac{\pi^2x}{2},
\label{FC0}
\end{eqnarray}
where $\Phi(r, s, z) = \sum_{n=0}^\infty \frac{r^n}{(n + z)^s}$ is the Lerch zeta function. These expressions are particularly useful for analyzing the system when one of the gate voltages is inactive, such as at $N_1 = 0$ or $N_2 = 0$, where the corresponding resonance width $\Gamma_j$ vanishes.

By substituting Eq. (\ref{eq:genG}) into Eq. (\ref{SCM}), we obtain the TP based on the GCM relation as follows
\begin{equation}
S_{GCM}\!=\!\frac{4\gamma}{3e}\sum_{j=1,2}\frac{1}{F_C}\frac{\partial F_C}{\partial p_j}\ln\left[\frac{E_{C,j}}{T+\Gamma_j}\right]|r_j|^2\sin\left(2\pi N_{j}\right).
\label{eq:genSCM}
\end{equation}
with
\begin{eqnarray}
\frac{\partial F_C}{\partial p_1}\!\!&=&\!\!\!\!\int_{-\infty}^{\infty}\!\!dx_1\int_{-\infty}^{\infty}\!\!dx_2\frac{p_2}{\left[x_2^2+p_2^2\right]}\frac{\left[x_1^2-p_1^2\right]}{\left[x_1^2+p_1^2\right]^2}\!\! \nonumber\\
&&\!\!\!\!\!\!\times\frac{(x_1-x_2)\left[(x_1-x_2)^2+4\pi^2\right]}{\sinh\left(\frac{x_1-x_2}{2}\right)\cosh\left(\frac{x_1}{2}\right)\cosh\left(\frac{x_2}{2}\right)},
\end{eqnarray}
and similar for $\partial F_C/\partial p_2$, where $p_1=\Gamma_1/T$ and $p_2=\Gamma_2/T$ with $\Gamma_j=\Gamma(|r_j|,N_j)$.

\begin{figure}[t]
\begin{center}
\includegraphics[scale=0.28]{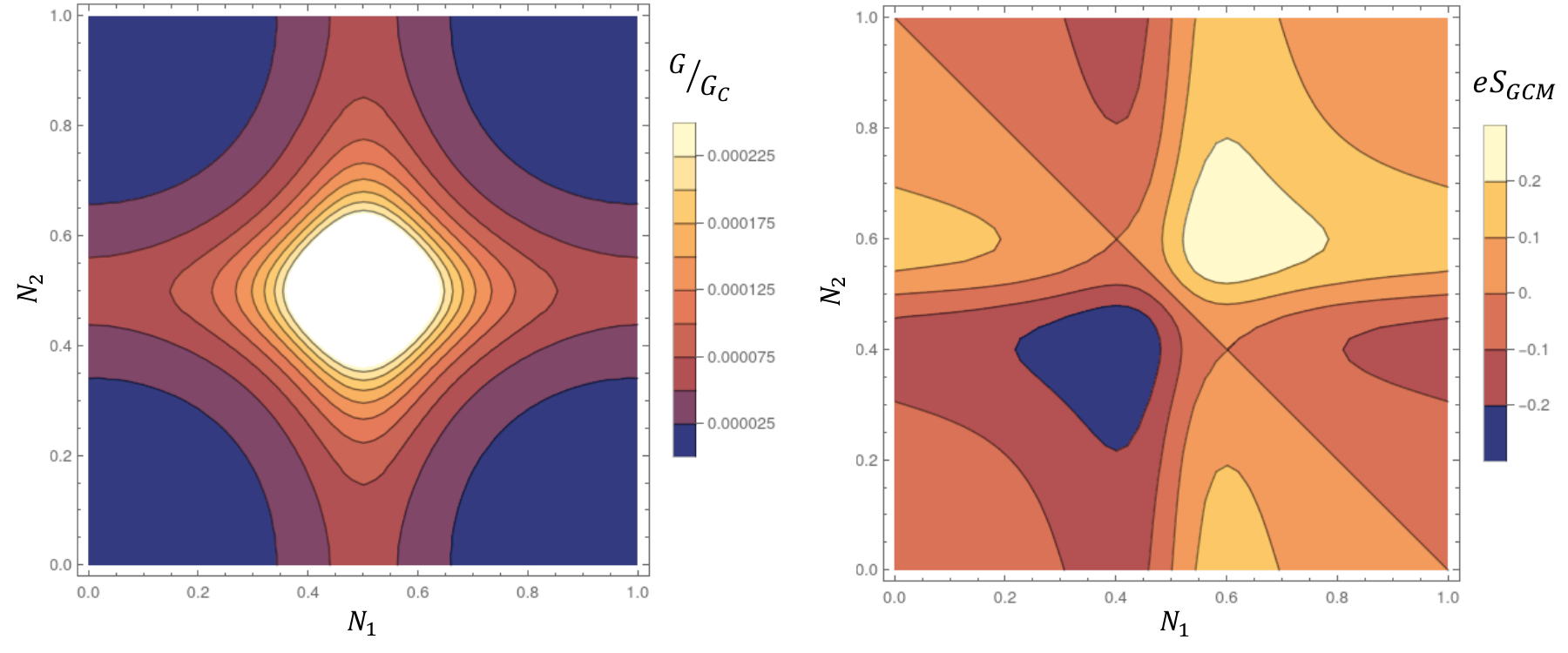}
\end{center}
\vspace{-0.2cm}
\caption{Contour plots of electric conductance $G/G_C$ and thermopower $eS_{GCM}$ computed from Eq. (\ref{eq:genSCM}) by applying GCM relation as functions of the dimensionless gate voltages $N_1$ and $N_2$ at temperature $T/E_C=0.01$. In both contour graphs, $|r_1|^2=|r_2|^2=0.1$, $E_{C,1}=E_{C,2}=E_C$.}
\label{GSGCMtem001}
\end{figure}

Equation (\ref{eq:genSCM}) represents the GCM we proposed for a 2SCKC. We validate this GCM relation by examining its dependence on gate voltages, as shown in Fig. \ref{GSGCMtem001}. While the electrical conductance is symmetric with respect to both $N_1$ and $N_2$, reaching its maximum at $N_1=N_2=0.5$, the TP is symmetric along the line $N_2=1-N_1$, where it vanishes and changes sign. Remarkably, the contour plot of TP calculated using Eq. (\ref{eq:genSCM}) as a function of $N_1$ and $N_2$ exhibits the same qualitative features as the TP obtained directly from $S=G_T/G$, which is shown in Fig.3 of Ref. \cite{thanh2024}.

We now examine the behavior of Eq.~(\ref{eq:genSCM}) in different limiting regimes, defined by the relative magnitudes of the energy scales ($\Gamma_1,\,\Gamma_2,\,T$).
\subsection{$\left(\Gamma_{1},\:\Gamma_{2}\right)\ll T$, non-Fermi liquid regime} \label{case1}

In the limit where both resonance widths are much smaller than the temperature, i.e., $\left(\Gamma_{1},\:\Gamma_{2}\right)\ll T$, the system enters the NFL regime. In this regime, the TP derived from the GCM formula, to leading (zeroth) order in $\Gamma_1/T,\, \Gamma_2/T$, is given by
\begin{equation}
S_{GCM}=-\frac{\mathcal{C}_{CM,1}}{e}\sum_{j=1,2}|r_j|^2\ln\left(\frac{E_{C,j}}{T}\right)\sin\left(2\pi N_{j}\right),
\label{SCMcase1}
\end{equation}
where $\mathcal{C}_{CM,1}=4\left(16\ln2-1\right)\gamma/3\left(8\pi-16\ln2+1\right)\approx 1.593$. In comparison, the TP directly calculated in Ref. \cite{thanh2024} exhibits weak NFL behavior at ``high'' temperature $T\gg\left(\Gamma_{1},\:\Gamma_{2}\right)$, and is given by
\begin{equation}
S=-\frac{\mathcal{C}_{dir,1}}{e}\sum_{j=1,2}|r_{j}|^{2}\ln\left(\frac{E_{C,j}}{T}\right)\sin\left(2\pi N_{j}\right),\label{eq:S_2NFL}
\end{equation}
where $\mathcal{C}_{dir,1}=9\gamma/8\approx 2$.

In terms of temperature scaling, Eq. (\ref{SCMcase1}) is consistent with both the perturbative result and the direct calculation in Eq. (\ref{eq:S_2NFL}). The proximity of the prefactors $\mathcal{C}_{CM,1}$ and $\mathcal{C}_{dir,1}$, suggests that the GCM relation remains valid in the NFL regime. 

\subsection{$\Gamma_{1}\ll T\ll\Gamma_{2}$,  non-Fermi liquid on the left and Fermi liquid on the right CKC}\label{case2}

This regime arises when the gate voltage $N_1$ is tuned closer to a Coulomb peak than $N_2$, so that the left quantum dot exhibits NFL behavior, while the right dot remains in the FL regime. The TPs obtained from the two approaches -- the GCM relation and direct calculation -- are given below:
\begin{eqnarray}
&&\!\!\!\!\!\!\!\!\!\!\!\!\!\!\!\!\!\! S_{GCM}=-\frac{\mathcal{C}_{CM,2}}{e}|r_1|^2\ln\left(\frac{E_{C,1}}{T}\right)\sin\left(2\pi N_{1}\right)\nonumber\\
&&\!\!\!\!\!\!\!\!\!\!\!\!\!\!\!\!\!\!-\frac{\pi^2}{3e}\frac{T}{E_{C,2}}\ln\left(\frac{\pi^2}{8\gamma|r_2|^2\cos^2\left(\pi N_2\right)}\right)\tan\left(\pi N_2\right),
\label{SCMcase2}
\end{eqnarray}
with $\mathcal{C}_{CM,2}=8\gamma\left[189\zeta(3)-68\right]/81\pi^3 \approx 0.903114$, where $\zeta(3)\approx 1.20206$ is the Riemann zeta function, which is related to the tetragamma function via $\Psi^{(2)}(1/2)=-14\zeta(3)$.
\begin{eqnarray}
&&\!\!\!\!\!\!\!\!\!\!\!\!\!\!\!\! S=-\frac{\mathcal{C}_{dir,2}}{e}|r_1|^2\ln\left(\frac{E_{C,1}}{T}\right)\sin\left(2\pi N_1\right)\nonumber\\
&&\!\!\!\!\!\!-\frac{\pi^3}{3e}\frac{T}{E_{C,2}}\ln\left(\frac{\pi^2}{8\gamma|r_2|^2\cos^2\left(\pi N_2\right)}\right)\tan\left(\pi N_2\right),
\label{Scase2}
\end{eqnarray}
where $\mathcal{C}_{dir,2}=1024\gamma/75\pi^2\approx 2.46$.

We find that the TPs in both Eqs.(\ref{SCMcase2}) and(\ref{Scase2}) consist of two distinct contributions: one reflecting the FL behavior of the right CKC, and the other capturing the NFL characteristics of the left subsystem. From Eq.(\ref{SCMcase2}), it is apparent that the original CM law applies well to the FL component but fails to accurately describe the NFL part. However, there exists an intermediate temperature regime, 
$T^*\ll T\ll \Gamma_{2}$ (where $T^*$ determined in Eq. (45) of Ref. \cite{thanh2024}), where the original CM relation remains valid. In contrast, by comparing Eqs.(\ref{Scase2}) and(\ref{SCMcase2}), it is clear that the GCM formula provides an accurate description for both components of the system, regardless of whether the underlying physics is FL or NFL.

\subsection{$\Gamma_{2}\ll T\ll\Gamma_{1}$, Fermi liquid on the left and non-Fermi liquid on the right CKC} \label{case3}

This case is the converse of case \ref{case2}, occurring when the gate voltage $N_2$ is tuned closer to the Coulomb peak than $N_1$. In this regime, the GCM formula yields the TP in the following form:
\begin{eqnarray}
&&\!\!\!\!\!\!\!\!\! S_{GCM}\!=\!-\frac{\pi^2}{3e}\frac{T}{E_{C,1}}\!\ln\!\left(\!\frac{\pi^2}{8\gamma|r_1|^2\cos^2\left(\pi N_1\!\right)}\!\right)\!\tan\left(\pi N_1\right)\nonumber\\
&&\;\;\;\;\;\;\;\; -\frac{\mathcal{C}_{CM,2}}{e}\ln\left(\frac{E_{C,2}}{T}\right)|r_2|^2\sin\left(2\pi N_2\right).
\label{SCMcase3}
\end{eqnarray}

The TP directly computed for this case also comprises two components: a FL characteristic on the left circuit and a NFL property on the right.
\begin{eqnarray}
\!\!\!S\!\!&=&\!\!\!-\frac{\pi^3}{3e}\frac{T}{E_{C,1}}\ln\left(\frac{E_{C,1}}{\Gamma_1}\right)\tan\left(\pi N_1\right)\nonumber\\
&&\!\!\!-\frac{\mathcal{C}_{dir,2}}{e}|r_2|^2\ln\left(\frac{E_{C,2}}{T}\right)\sin\left(2\pi N_2\right).
\end{eqnarray}

In this limit, the GCM relation remains applicable to both parts of the system. In contrast, the original CM relation fails to describe the right part, which exhibits NFL behavior, but still accurately captures the FL characteristics of the left part. Notably, the CM relation holds within the temperature window $T^{**}\ll T\ll \Gamma_{1}$, where $T^{**}$ is defined in Eq.(47) of Ref.\cite{thanh2024}.

\subsection{$T\ll\left(\Gamma_{1},\:\Gamma_{2}\right)$,  Fermi-liquid regime} \label{case4}

This limit corresponds to the infinitesimally low temperature regime, which is inaccessible to perturbative approaches. The TPs obtained from both methods exhibit a linear temperature dependence and can be expressed analytically as
\begin{eqnarray}
\!\!\!\!\!\!\!\!\! S_{GCM}\!\!\!\!&=&\!\!\!-\frac{\pi^2}{3e}\!\!\!\sum_{j=1,2}\!\!\!\frac{T}{E_{C,j}}\!\ln\!\!\left[\!\frac{\pi^2}{8\gamma|r_j|^2\!\cos^2\left(\!\pi N_j\!\right)}\!\right]\!\!\tan\!\left(\pi N_j\!\right)\!\!,
\label{SCMcase4}
\end{eqnarray}
and 
\begin{eqnarray}
\!\!\!\!\!\! S \!\!\!&=&\!\!\!\! -\frac{3\pi^3}{7e}\!\!\!\sum_{j=1,2}\!\!\frac{T}{E_{C,j}}\!\ln\!\!\left[\!\frac{\pi^2}{8\gamma|r_j|^2\!\cos^2\left(\!\pi N_j\!\right)}\!\right]\!\tan\!\left(\pi N_{j}\!\right)\!.
\label{Scase4}
\end{eqnarray}
Similarly to the regimes discussed above, the close agreement between Eqs.(\ref{SCMcase4}) and (\ref{Scase4}) confirms the validity of the GCM theory in the full FL regime. Our proposed GCM formula [Eq.(\ref{SCM})] not only remains consistent with Eq.(4) in Ref.\cite{MAprl}, but also provides deeper insight into the thermoelectric response of strongly correlated nanostructures.

The results of the four different regimes demonstrate that the GCM formula serves as a standard benchmark for the 2SCK model, regardless of whether the system is in the FL or NFL state \cite{MAprl, MAprb, thanh2010, thanh2015}. Furthermore, the application of the GCM formula across these four regimes reinforces our confidence in the advantages of a nonperturbative approach to the two-channel charge Kondo problem. 

\begin{figure}[t]
\begin{center}
\includegraphics[scale=0.24]{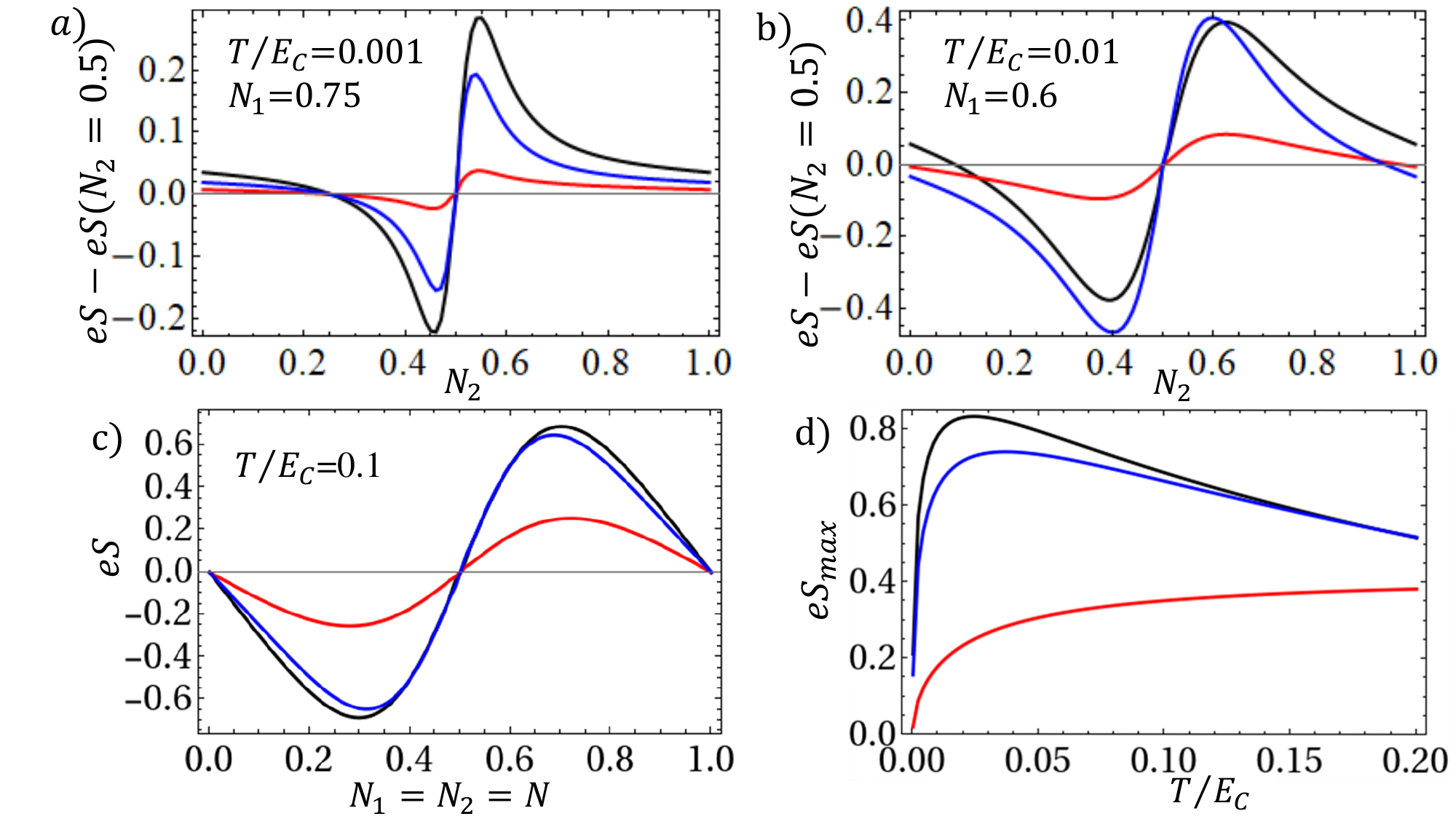}
\end{center}
\vspace{-0.5cm}
\caption{Thermopower $eS$ as a function of the gate voltage $N_2$ at different temperatures a) $T/E_C=0.001$, $N_1=0.75$, b) $T/E_C=0.01$, $N_1=0.6$ and c) $T/E_C=0.1$, $N_1=N_2=N$. Panel d) shows the maximum of thermopower as a function of temperature $T/E_C$. In all graphs, $|r_1|^2=|r_2|^2=0.1$. The black lines are directly plotted (as in Ref. \cite{thanh2024}), the red lines are plotted from the original Cutler-Mott formula [see Eq. (\ref{SCM0})] multiplied by a factor of 2.22, and the blue lines are plotted from the generalized Cutler-Mott relation [see Eq. (\ref{eq:genSCM})], also multiplied by the same factor to match the red line in each panel.}
\label{SCM2CKC}
\end{figure}

We now examine Eq.(\ref{eq:genSCM}) numerically. The comparison reveals a strong consistency between the GCM formula and the direct calculation, which differs only by a constant prefactor of approximately $2.22$. This agreement is illustrated in Fig.\ref{SCM2CKC}, where the blue curves - generated using the GCM relation [Eq.(\ref{eq:genSCM})] - closely follow the behavior of the black curves, which are obtained from direct evaluations of Eqs.(\ref{eq:genG}) and~(\ref{eq:genGT}). The presence of the logarithmic term in the GCM expression effectively captures the hallmark features of Kondo correlations and reinforces the interpretation of NFL behavior.

\section{Application of the generelized Cutle-Mott relation to some charge Kondo setups}

In this section, we apply the GCM relation to two distinct charge Kondo implementations that are related to the 2SCK system. These setups provide practical platforms to test and illustrate the relevance of the GCM framework in experimentally motivated configurations.

\subsection{Effect of one gate voltage is inactive: either $N_1=0.5$ or $N_2=0.5$}

In the limit $N_1=0.5$, $N_2=N$, the TP is calculated from GCM as shown in Eq. (\ref{eq:genSCM}), is 
\begin{equation}
S_{GCM}=\frac{4\gamma}{3e}\frac{\partial_p F_{C,0}(p)}{F_{C,0}(p)}|r|^2\ln\left[\frac{E_{C}}{T+\Gamma}\right]\sin(2\pi N),
\label{SCM_1CKC}
\end{equation}
with $F_{C,0}(p)$ is defined in Eq. (\ref{FC0}). The TP as a function of gate voltage in this limit can be seen in the right panel of Fig. \ref{GSGCMtem001}. At exact Coulomb peaks (i.e., when either $N_1=0.5$ or $N_2=0.5$), the particle-hole symmetry is preserved, irrespective of the scattering at the QPCs. As a result, the contribution from the corresponding segment of the system to the TP vanishes.

Interestingly, either $N_1=0.5$ or $N_2=0.5$ situation reduces to the case similar to the Matveev-Andreev (MA) setup \cite{MAprb}, in which the left (right) structure simply is a normal metal lead, the electric conductance in Eq. (\ref{eq:genG}) becomes $G=(G_C T/8\gamma E_C)\int(x^2+\pi^2)/\cosh(x/2)[x^2+(\Gamma/T)^2]$ and the Eq. (\ref{eq:genSCM}) induces
\begin{equation}
S_{GCM}=\frac{4\gamma}{3e}\ln\left[\frac{E_{C}}{T+\Gamma}\right]\frac{\partial_p F_0(p)}{F_0(p)}|r|^2\sin(2\pi N),
\label{SCM_MA_1CKC}
\end{equation}
with \cite{Kiselev2025}
\begin{eqnarray}
&& \!\!\!\!\!\!\!\!\!\!\!\!\!\!\!\!\!\!\!\!F_{0}\left(p\right)=\int_{-\infty}^{\infty}dx\frac{x^2+\pi^2}{\cosh^2(x/2)}\frac{p}{x^2+p^2}\nonumber\\
&&\!\!\!\!\!\!\!\!\!\!\!\!\!\!\!\!\!\!\!\!=2\pi\Psi^{(1)}\left(\frac{1}{2}+\frac{p}{2\pi}\right)+4p\left[1-\frac{p}{2\pi}\Psi^{(1)}\left(\frac{1}{2}+\frac{p}{2\pi}\right)\right],\label{F0}
\end{eqnarray}
with $\Psi^{(1)}(x)=\sum_{n=0}^\infty(x+n)^{-2}$ is the trigamma function. So  
\begin{eqnarray}
&&\partial_p F_0(p)=4-\frac{4p}{\pi}\Psi^{(1)}\left(\frac{1}{2}+\frac{p}{2\pi}\right)\nonumber\\
&&+\left[1-\frac{p^2}{\pi^2}\right]\Psi^{(2)}\left(\frac{1}{2}+\frac{p}{2\pi}\right),\label{dF0}
\end{eqnarray}
where $\Psi^{(2)}(x)=\partial_x \Psi^{(1)}(x)$, being the tetragamma function. Replacing Eqs. (\ref{F0}) and (\ref{dF0}) into Eq. (\ref{SCM_MA_1CKC}), we obtain the generalized formula of Eq. (32) in Ref. \cite{Karki2020} with added prefactor $\ln[E_{C}/(T+\Gamma)]$.

In the limit $T\ll\Gamma$, we obtain:
\begin{equation}
S_{GCM}=-\frac{\pi^2}{3e}\ln\left[\frac{E_{C}}{\Gamma}\right]\frac{T}{E_C}\tan(\pi N).
\label{SCM_MA_1CKC_lim1}
\end{equation}
The result is in agreement with that reported in Ref.\cite{MAprl}, aside from a discrepancy in the prefactor, which is $-5/3\pi$. Remarkably, the coefficient presented in Eq.(4) of Ref.~\cite{MAprl} deviates from that in the original CM relation.

\begin{figure}[t]
\begin{center}
\includegraphics[scale=0.25]{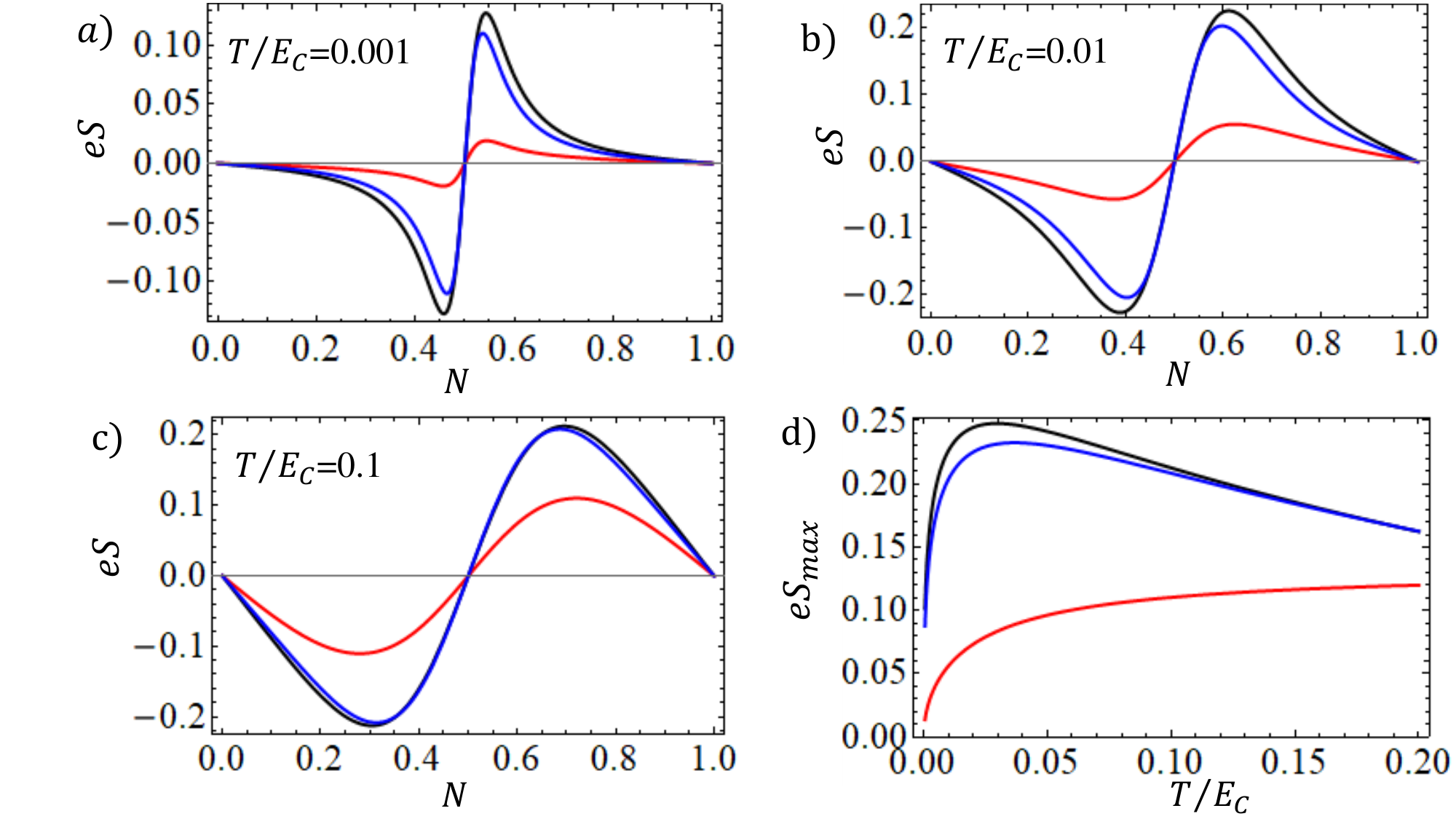}
\end{center}
\vspace{-0.5cm}
\caption{Thermopower $eS$ as a function of the gate voltage $N$ at different temperatures: a) $T/E_C=0.001$, b) $T/E_C=0.01$, and c) $T/E_C=0.1$. Panel d) shows the maximum of thermopower as a function of temperature $T/E_C$. In all graphs: $|r|^2=0.1$. The black lines are plotted directly (as in Refs. \cite{MAprl, MAprb}), the red lines are plotted from the original Cutler-Mott formula [see Eq. (\ref{SCM0})], multiplied by a factor of $1.35$, and the blue lines are plotted from the generalized Cutler-Mott relation [see Eq. (\ref{SCM_MA_1CKC})], also multiplied by a factor of $1.35$.}
\label{SCM1CKC}
\end{figure}

In the opposite limit $T\gg \Gamma$, from Eq. (\ref{SCM_MA_1CKC}), we obtain the TP for single-site charge Kondo as 
\begin{equation}
S_{GCM}=\frac{8\gamma\left[2-7\zeta(3)\right]}{3\pi^3e}\ln\left[\frac{E_{C}}{T}\right]|r|^2\sin(2\pi N).
\label{SCM_1CKCper}
\end{equation}
Indeed, Eq. (\ref{SCM_1CKCper}) is similar to the Eq. (51) in Ref. \cite{MAprb} except the constant prefactor. It covers the perturbative result. A general situation is shown in Fig. \ref{SCM1CKC}. We find that the GCM formula multiple of a factor $1.35$ is stunningly consistent with the result from direct computation in Refs. \cite{MAprl,MAprb}.

\subsection{Strongly asymmetric reflection amplitudes $|r_1|^2\ll|r_2|^2$ in the weak coupling between single- and two-channel charge Kondo circuits}

\begin{figure}[t]
\begin{center}
\includegraphics[scale=0.25]{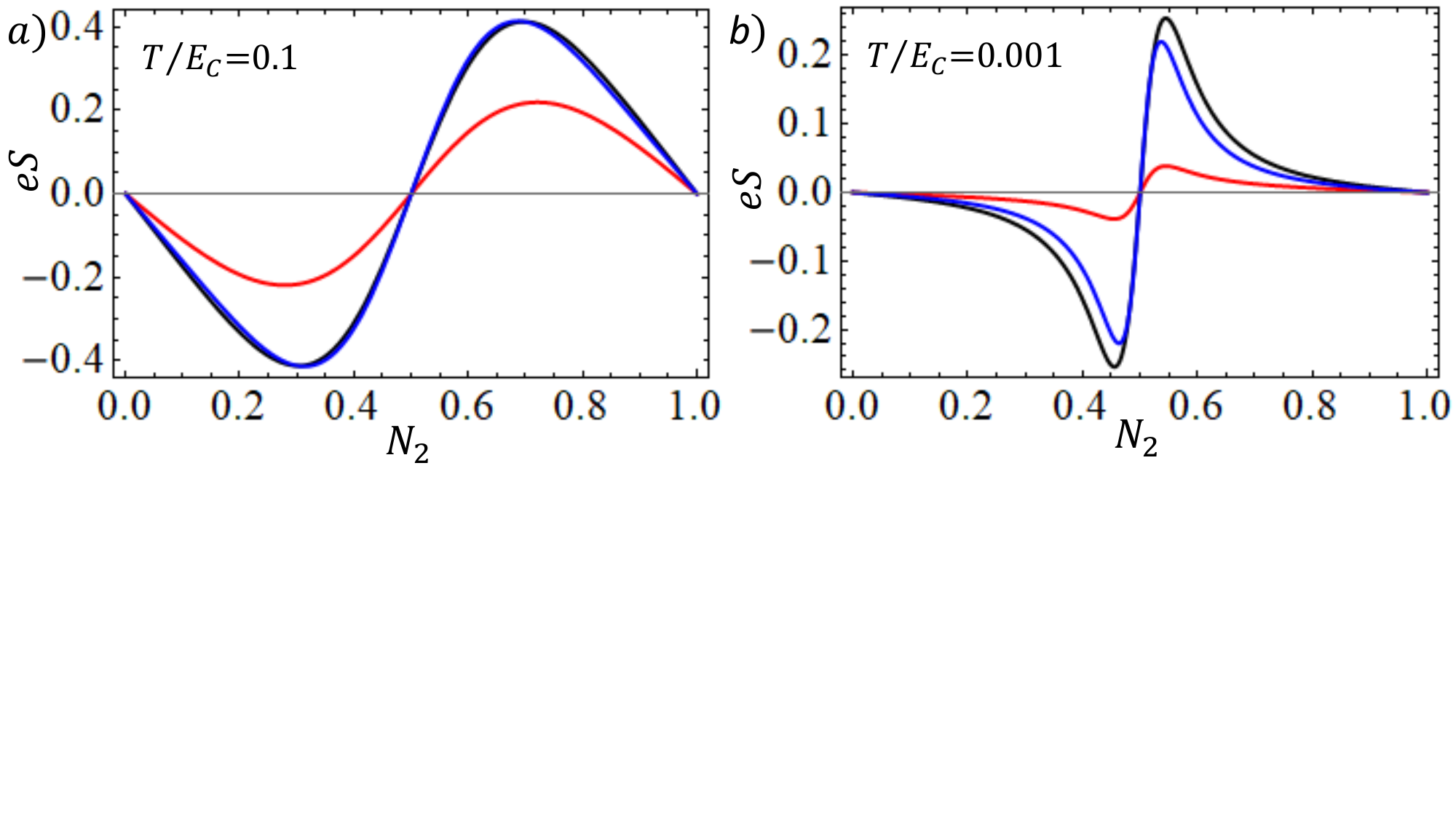}
\end{center}
\vspace{-2cm}
\caption{Thermopower $eS$ as a function of the gate voltage $N_2$ at different temperatures: a) $T/E_C=0.1$ and b) $T/E_C=0.001$. In both graphs, $|r_2|^2=0.1$. The black lines are plotted directly [see Eq. (\ref{S1CK2CK})], the red lines are plotted from the original Cutler-Mott formula, multiplied by a factor of $2.8$, and the blue lines are plotted from the generalized Cutler-Mott relation [see Eq. (\ref{SCM1CK2CK})], also multiplied by the same factor of $2.8$.}
\label{SCM1CK2CKasymmetry}
\end{figure}

The description of the weakly coupled single- and two-channel Kondo simulators involves the left CKC being in the FL-1CK state and the right CKC operating in the NFL-2CK state. According to Ref. \cite{thanh2024}, when $|r_1|^2\ll|r_2|^2$, the TP is 
\begin{equation}
S=-\frac{12\gamma}{5\pi e}\frac{F_T(p_2)}{F_G(p_2)}|r_2|^2\ln\left(\frac{E_{C,2}}{T+\Gamma_2}\right)\sin(2\pi N_2),
\label{S1CK2CK}
\end{equation}
with \cite{Kiselev2025}
\begin{eqnarray}
&& \!\!\!\!\!\!\!\!\!\!\!\!\!\!\!\!\!\!\!\!\!\! F_G(p_2)\!=\!\!\int_{-\infty}^{\infty} \!\!\frac{p_2\left(u^2+\pi^2\right)\left(u^2+9\pi^2\right)}{\cosh^2\left(\frac{u}{2}\right)\left(u^2+p_2^2\right)}du=-4p_2^3\nonumber\\
&&\!\!\!\!\!\!\!\!\!\!\!\!\!\!\!\!\!\!\!\!\!\!+\frac{124\pi^2}{3}p_2+\frac{2}{\pi}\left(p_2^4-10\pi^2p_2^2+9\pi^4\right)\Psi^{(1)}\left(\frac{1}{2}+\frac{p_2}{2\pi}\right),
\label{FGa}
\end{eqnarray}
and \cite{Kiselev2025}
\begin{eqnarray}
&&\!\!\!\!\!\!\!\!\!\!\!\!\! F_T(p_2)=\int_{-\infty}^{\infty} \frac{ u^2\left(u^2+\pi^2\right)\left(u^2+9\pi^2\right)}{\cosh^2\left(\frac{u}{2}\right)\left(u^2+p_2^2\right)}\nonumber\\
&&\!\!\!\!\!\!\!\!\!\!\!\!\!=4p_2^4-\frac{124\pi^2}{3}p_2^2+\frac{256\pi^4}{5}\nonumber\\
&&\!\!\!\!\!\!\!\!\!\!\!\!\!-\frac{2}{\pi}\left(p_2^5-10\pi^2p_2^3+9\pi^4\right)\Psi^{(1)}\left(\frac{1}{2}+\frac{p_2}{2\pi}\right),
\label{FTa}
\end{eqnarray}
The TP obtained from GCM formula (Eq. \ref{eq:genSCM}) has form 
\begin{equation}
S_{GCM}=\frac{4\gamma}{3e}\frac{\partial_{p_2}F_G}{F_G}\ln\left(\frac{E_{C,2}}{T+\Gamma_2}\right)|r_2|^2\sin(2\pi N_2).
\label{SCM1CK2CK}
\end{equation}
We demonstrate the consistence of the GCM formula with the direct calculation in Fig. \ref{SCM1CK2CKasymmetry}. 

\section{Generalized Cutler-Mott relation for the figure of merit $ZT$}

The figure of merit, denoted as $ZT$, is a critical parameter in evaluating the performance of thermoelectric materials. It combines the material's thermoelectric efficiency by incorporating its electric conductance $G$, thermal conductance $\mathcal{K}$, and TP $S$ into a single dimensionless quantity. The figure of merit is defined as
\begin{equation}
ZT=\frac{S^2G T}{\mathcal{K}}; \;\; \mathcal{K}=\left.\frac{\partial I_h}{\partial\Delta T}\right|_{I_e=0}=G_H-T\frac{G_T^2}{G}.
\label{ZTgen}
\end{equation}
A high $ZT$ indicates a material's potential for efficient thermoelectric conversion, with high values of $ZT$ leading to better performance in applications such as power generation and cooling. The goal in thermoelectric research is to maximize $ZT$, which requires optimizing the balance between electrical and thermal conductivity. While high electrical conductance is desirable for efficient charge transport, low thermal conductance is equally important to prevent the loss of heat. Achieving high thermoelectric efficiency is thus a complex challenge that requires careful material design, particularly in advanced materials with low-dimensional or nanostructured properties. As such, the figure of merit plays a pivotal role in guiding the development of next-generation thermoelectric materials that could significantly improve energy conversion efficiency in practical devices.

Interestingly, the ratio of the electronic thermal conductance $\mathcal{K}$ and the electrical conductance $G$ is stated in the WF law. It is known that in the low temperature regime of a macroscopic sample, this ratio is proportional to the temperature $T$, with the proportionality constant being the Lorenz number $L_0$ \cite{Benenti,Zlatic}:
\begin{equation}
\frac{\mathcal{K}}{G}=L_0 T,
\end{equation}
where $L_0=\pi^2/3$. Although transport in nanodevices is generally expected to deviate from the WF law, even within the FL regime \cite{Benenti}, recent reports suggest that the WF law holds even in the NFL regime of Kondo effects \cite{Mitchell, KarkiKiselev2020}. Moreover, the universal value of the Lorenz ratio $R\equiv L/L_0=\left(3/\pi^2\right)\left(\mathcal{K}/GT\right)$ exhibits charge Kondo correlations \cite{Karki2020}. For the 2SCK, the Lorenz ratio satisfies \cite{Kiselev2023}
\begin{eqnarray}
R&=&\frac{12}{5}-\frac{3}{\pi^2}S^2.
\end{eqnarray}
As we find in the previous sections, $S\ll 1$, we can have 
\begin{equation}
ZT\approx \frac{5}{4\pi^2}S^2.
\label{ZT1}
\end{equation}
Therefore, we can get the generalized Cutler-Mott relation for the figure of merit as 
\begin{equation}
\!\! ZT_{GCM}\!=\!\frac{5\pi^2}{144e^2}\!\!\left[\sum_{j=1,2}\!\frac{T}{E_{C,j}}\!\ln\!\left(\!\frac{E_{C,j}}{T+\Gamma_j}\!\right)\!\frac{\partial\ln G}{\partial N_j}\!\right]^2\!\!,
\label{ZTGCM}
\end{equation}
or
\begin{eqnarray}
&&\!\!\!\!\! ZT_{GCM}\approx\frac{20\gamma^2}{9e^2\pi^2}\nonumber\\
&&\!\!\!\!\!\times\!\!\left[\sum_{j=1,2}\!\!\frac{1}{F_C}\frac{\partial F_C}{\partial p_j}\ln\left(\frac{E_{C,j}}{T+\Gamma_j}\right)|r_j|^2\sin\left(2\pi N_{j}\right)\!\right]^2\!\!\!.
\label{genZTCM}
\end{eqnarray}
\begin{figure}[t]
\begin{center}
\hspace{-0.65cm}\includegraphics[scale=0.265]{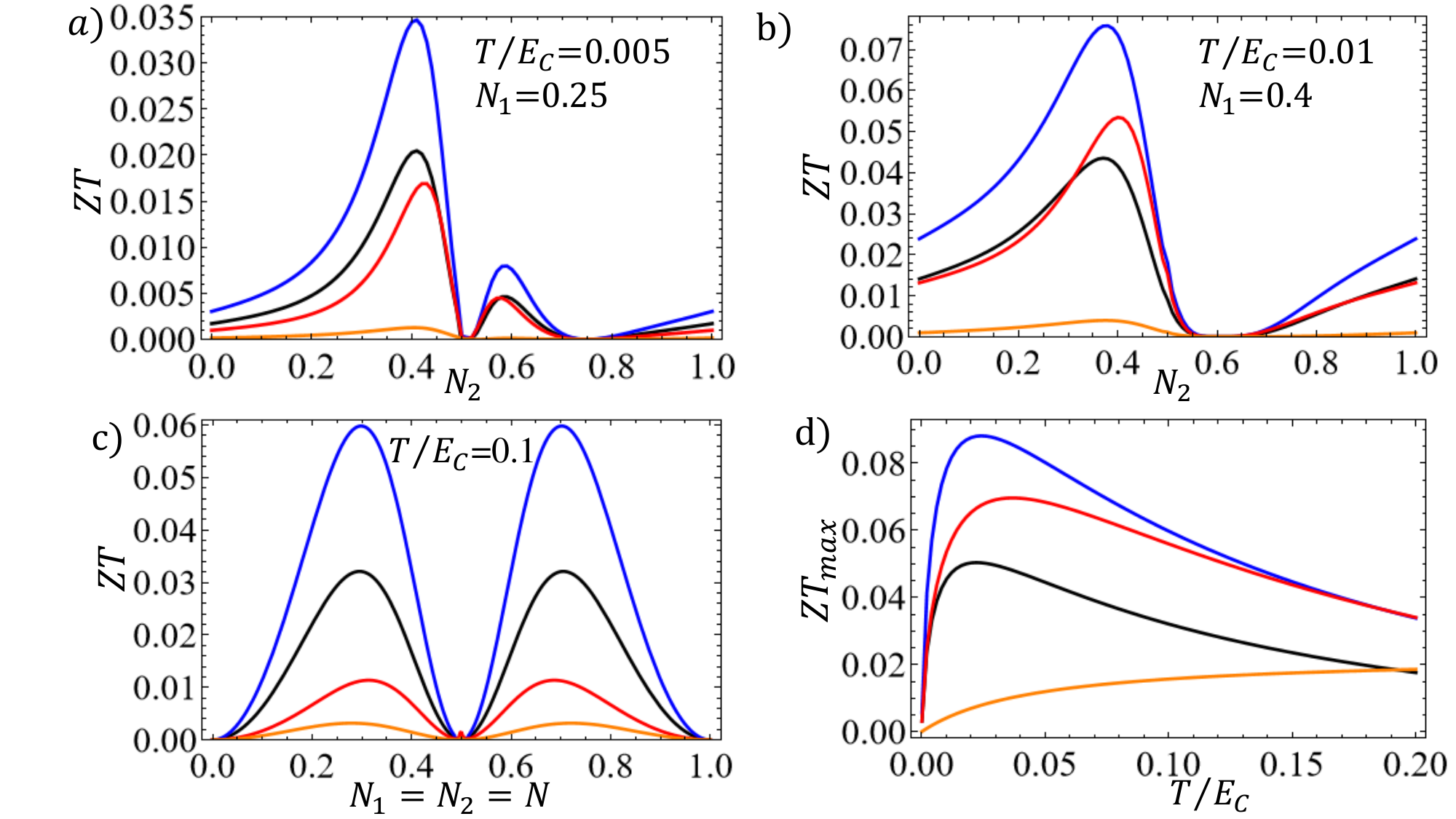}
\end{center}
\vspace{-0.5cm}
\caption{Figure of merit $ZT$ as a function of the gate voltage $N_2$ at different temperatures a) $T/E_C=0.005$, $N_1=0.25$, b) $T/E_C=0.01$, $N_1=0.4$ and c) $T/E_C=0.1$, $N_1=N_2=N$. Panel d) shows the maximum of figure of merit $ZT_{max}$ as a function of temperature $T/E_C$. In all graphs, $|r_1|^2=|r_2|^2=0.1$. The black lines are directly plotted [as in Eq.(\ref{ZTgen})], the blue lines are plotted by approximated computation [as in Eq.(\ref{ZT1})], the orange lines are plotted from the original CM formula multiplied by a factor of $4.93$, and the blue lines are plotted from the GCM relation [see Eqs. (\ref{ZTGCM}, \ref{genZTCM})], also multiplied by the same factor to match the orange line in each panel.}
\label{ZTCM2CKC}
\end{figure}

The GCM relation for figure of merit in Eq. (\ref{ZTGCM}) is now examined numerically. In Fig. \ref{ZTCM2CKC}, we show the plots of figure of merit obtained from direct calculation in Eq.(\ref{ZTgen}) [black lines, with $G$, $G_T$, $G_H$ are taken from general formulas (\ref{eq:genG}), (\ref{eq:genGT}), and (\ref{eq:genGH}), approximated computation in Eq.(\ref{ZT1}) (blue lines), CM relation multiplied by a factor of $4.93$ (orange lines), and GCM relation in Eqs. (\ref{ZTGCM}) and (\ref{genZTCM}) (red lines). The most important plot is the maximum value of $ZT$ as a function of temperature $T/E_C$ [see panel d)] in which the consistency between the GCM formula and the direct calculation (the difference is only by a factor of $4.93$. Notably, the red lines exhibit the same behavior as the black and blue lines (obtained from direct calculations in Eqs. (\ref{ZTgen}) and (\ref{ZT1})). The logarithmic term in the GCM expression captures the essence of Kondo correlations and supports the NFL interpretation.

\section{Conclusion}

In summary, we have calculated the TP by applying our proposed GCM formula to a quantum circuit with weak coupling between two QD-QPC structures, each corresponding to a charge Kondo simulator. For comparison, we also present the TP formulas obtained through direct computation in Ref. \cite{thanh2024}.

We apply a nonperturbative treatment, which not only encompasses the perturbative results but also extends to smaller temperature regimes, $T\ll\min[|r_j|^2E_{C,j}]$. While previous works \cite{MAprl, Karki2020} have shown that the CM formula is applicable when the system is in the FL regime ($T\ll \min[|r_j|^2E_{C,j}]$) , with deviations appearing when the system enters the NFL state ($T\gg \max[|r_j|^2E_{C,j}]$),we propose the GCM formula for computing the TP in a 2SCK setup. We investigate the GCM formula for various configurations, including 2CK-2CK, strong asymmetric 1CK-2CK, and one-site models. Our results show that the GCM relation holds across all scenarios.

The GCM formula provides a reliable approximation at both low and high temperatures. Comparing the TP formulas obtained using the GCM formula and direct calculations is crucial, as it enables the determination of TP from electric conductance measurements. At low temperatures ($T\ll \min[|r_j|^2E_{C,j}]$), the GCM law serves as a criterion for FL properties, similar to the original CM formula. At higher temperatures ($T\gg \max[|r_j|^2E_{C,j}]$), when the system enters the NFL state, the GCM relation reflects the effects of strong iso-spin correlations via the term $\ln[E_{C,j}/T]$.

Furthermore, our proposed GCM formula can be applied for the figure of merit $ZT$. The general approach for deriving the GCM is to incorporate modification terms of the form $\{\ln[E_C/(T+\Gamma_{j})]\}^2$ for each $\partial F_{C}/\partial \Gamma_{j}$, where $j$ indexes the number of Kondo resonance widths in the charge Kondo system.

\section*{Acknowledgements}

This research in Hanoi is funded by Vietnam National Foundation for Science and Technology Development (NAFOSTED) under grant number 103.01-2023.03. T.K.T.N. would like to acknowledge support from the ICTP through the Associates Programme (2024-2029).  The work of M.N.K. is conducted within the
framework of the Trieste Institute for Theoretical Quantum Technologies (TQT).

\appendix

\section{Previous Results}
\renewcommand{\theequation}{A.\arabic{equation}}
\setcounter{equation}{0}
\label{previous}

To study the thermoelectric effects at the weak link between two QDs in the linear response regime, where $[\Delta T, e\Delta V] \ll T$, we build on the theoretical framework developed in Refs.~\cite{thanh2018,thanh2024,thanhcom2022}. This approach evaluates both the charge current $I_e$ and heat current $I_h$ through the tunnel contact based on the Onsager reciprocity relations~\cite{Onsager}.

Central to this analysis is the local density of states (LDOS) of the QDs at the weak link, which are expressed using the correlation function $K_j(1/2T + it)$. This function captures the effects of Coulomb interactions and is derived following the Matveev-Andreev theory~\cite{MAprl,MAprb}. Near the Coulomb blockade peaks, the Kondo-resonance width $\Gamma_j$ plays a significant role and is given by:
\begin{eqnarray}
\Gamma_{j}\left(N_{j}\right)=\frac{8\gamma E_{C,j}}{\pi^{2}}|r_{j}|^{2}\cos^{2}(\pi N_{j}).\label{eq:Gam}
\end{eqnarray}
The leading-order expression for the correlation function is
\begin{eqnarray}
& & K_{j}\!\left(\frac{1}{2T}+it\right)=\frac{\pi T\Gamma_{j}}{\gamma E_{C,j}}\frac{1}{\cosh(\pi Tt)}\nonumber \\
 &  & \times\int_{-\infty}^{\infty}\frac{e^{\omega\left(1/2T+it\right)}}{\left(\omega^{2}+\Gamma_{j}^{2}\right)\left(1+e^{\omega/T}\right)}d\omega\nonumber\\
& &\!-\frac{4T}{E_{C,j}}\frac{|r_{j}|^{2}\sin\left(2\pi N_{j}\right)}{\cosh(\pi Tt)}\ln\left(\frac{E_{C,j}}{T+\Gamma_{j}}\right)\nonumber \\
 &  & \times\int_{-\infty}^{\infty}\frac{\omega e^{\omega\left(1/2T+it\right)}}{\left(\omega^{2}+\Gamma_{j}^{2}\right)\left(1+e^{\nicefrac{\omega}{T}}\right)} d\omega~. 
\label{eq:Knonper}
\end{eqnarray}

Using these expressions, the transport coefficients \cite{Onsager} in the linear response regime can be derived as follows \cite{thanh2018,thanh2024,thanhcom2022} (in the units $\hbar$$=$$c$$=$$k_B$$=$$1$).

The electric conductance $G=\left.\partial I_e/\partial\Delta V\right|_{\Delta T=0}$ is obtained as:
\begin{equation}
G=\frac{G_C}{24\gamma^2}\frac{T^2}{E_{C,1}E_{C,2}}F_C\!\!\left(\!\frac{\Gamma_{1}}{T},\frac{\Gamma_{2}}{T}\!\right)\!,\label{eq:genG}
\end{equation}
where $G_{C}=2\pi e^2 \nu_{0,1}\nu_{0,2}|t|^2$ is a conductance of the central (tunnel) area, which is characterized by the tunneling amplitude $|t|$, local densities of states $\nu_{0,1},\,\nu_{0,2}$  assuming that the electrons in the QDs are non-interacting.
\begin{equation}
F_{C}\left(p_{1},p_{2}\right)=\!\!\int_{-\infty}^{\infty}\!\!\!\!\!\! dz\int_{-\infty}^{\infty}\!\!\!\!\!\! du\,F\left(p_{1},p_{2},z,u\right),\label{FC}
\end{equation}
in which $p_1=\Gamma_1/T$ and $p_2=\Gamma_2/T$.
\begin{eqnarray}
&&F\left(p_{1},p_{2},z,u\right)=\frac{p_{1}p_{2}u\left[u^{2}+4\pi^{2}\right]}{\sinh\left(\frac{u}{2}\right)\left[\cosh\left(z\right)+\cosh\left(\frac{u}{2}\right)\right]}\nonumber \\
&&\times\frac{1}{\left[\left(z+\frac{u}{2}\right)^{2}+p_{1}^{2}\right]\left[\left(z-\frac{u}{2}\right)^{2}+p_{2}^{2}\right]}.\label{FFF}
\end{eqnarray}

The thermoelectric coefficient $G_T=\left.\partial I_e/\partial\Delta T\right|_{\Delta V=0} $ is:
\begin{eqnarray}
\!\!\!\! &  & \!\!\!\!G_{T}=-\frac{G_C}{6 e\gamma\pi}\frac{T^3}{E_{C,1}E_{C,2}}\nonumber \\
\!\!\!\! &  & \!\!\!\!\!\!\!\!\times\left\{ \frac{|r_{1}|^{2}}{\Gamma_{1}}\ln\left(\frac{E_{C,1}}{T+\Gamma_{1}}\right)\!\sin\left(2\pi N_{1}\right)F_{T,s}\!\left(\frac{\Gamma_{1}}{T},\frac{\Gamma_{2}}{T}\right)\right.\nonumber \\
\!\!\!\! &  & \!\!\!\!\!\!\!+\!\left.\frac{|r_{2}|^{2}}{\Gamma_{2}}\!\ln\left(\!\frac{E_{C,2}}{T+\Gamma_{2}}\!\right)\!\sin\left(2\pi N_{2}\right)F_{T,m}\!\left(\!\frac{\Gamma_{1}}{T},\frac{\Gamma_{2}}{T}\!\right)\!\right\}\! ,\label{eq:genGT}
\end{eqnarray}
where 
\begin{equation}
\!\!\!\!F_{T,s}\!\left(p_{1},p_{2}\right)=\!\!\int_{-\infty}^{\infty}\!\!\!\!\!\!dz\!\!\int_{-\infty}^{\infty}\!\!\!\!\!\!du\left(z+\frac{u}{2}\right)zF\left(p_{1},p_{2},z,u\right),\label{FTC1}
\end{equation}
\begin{align}
\!\!\!\!F_{T,m}\!\left(p_{1},p_{2}\right)=\!\!\!\int_{-\infty}^{\infty}\!\!\!\!\!\!dz\!\!\int_{-\infty}^{\infty}\!\!\!\!\!\!du\left(z-\frac{u}{2}\right)zF\left(p_{1},p_{2},z,u\right).\label{FTC2}
\end{align}

The thermal coefficient (or heat current response)  $G_H=\left.\partial I_h/\partial\Delta T \right|_{\Delta V=0}$  is obtained as 
\begin{eqnarray}
&&\!\!\!\!\!\! G_H\!=\!\!\frac{G_C}{240\gamma^2e^2}\frac{T^3}{E_{C,1}E_{C,2}}F_H\!\!\left(\!\frac{\Gamma_{1}}{T},\frac{\Gamma_{2}}{T}\!\!\right)\!,
\label{eq:genGH}
\end{eqnarray}
where 
\begin{equation}
\!\! F_H\!\left(p_{1},p_{2}\right)\!=\!\!\!\int_{-\infty}^{\infty}\!\!\!\!\!\!\!dz\!\int_{-\infty}^{\infty}\!\!\!\!\!\!\!du\left(-9u^2+20z^2+16\pi^2\right)\!F\!\left(p_{1},p_{2},z,u\right).
\label{FH}
\end{equation}

These analytical results form the basis for evaluating the transport properties in the current study, where we further explore the expressions for TP and the figure of merit from the CM formula.


\end{document}